\documentclass[%
 reprint,
superscriptaddress,
 amsmath,amssymb,
 aps,
pra,
longbibliography
]{revtex4-2}

\usepackage{graphicx}% Include figure files
\usepackage{dcolumn}% Align table columns on decimal point
\usepackage{bm}% bold math
\usepackage{hyperref}% add hypertext capabilities

\usepackage{cleveref}
\usepackage{comment}
\Crefname{equation}{Eq.}{Eqs.}

\usepackage{amsthm}
\usepackage{thmtools}

\usepackage{color}

\usepackage[normalem]{ulem}
\DeclareRobustCommand\mpwS[1]{{\let\helpcmd\@firstofone\parhelp#1\par\relax\relax} }
\long\def\parhelp#1\par#2\relax{%
	\helpcmd{#1}\ifx\relax#2\else\par\parhelp#2\relax\fi%
}

\newcommand{\mpwh}[1]{{\unskip}}%\hspace{-0.17cm}}}

\begin{document}

\preprint{APS/123-QED}

\title{Characterising the precision of a clock without any external time reference}

\author{Pablo Á. Domínguez}
 \affiliation{University Grenoble Alpes, Grenoble, France}
 \affiliation{Institute for Theoretical Physics, ETH Zürich, Switzerland}

\author{Christopher Chubb}
\affiliation{Independent Researcher}

\author{Mischa P. Woods}
\affiliation{ENS Lyon, Inria, France}
\affiliation{University Grenoble Alpes, Inria, Grenoble, France}



\date{\today}% It is always \today, today,

% Include institute/department
% Eu: ETHZ, UGA, (Inria?)
% Christopher: ETHZ
% Mischa: Inria, UGA, ENS Lyon

\begin{abstract}
Determining a clock's precision usually requires comparison with a better time reference, yet no ideal reference is physically attainable. We show how a single clock can characterise its own precision using only a signal splitter and delayed trigger, whose uncertainty is explicitly accounted for. The resulting estimator depends only on tick-count statistics. A purpose-built electronic circuit validates the method without an external timekeeper or a second clock.
\end{abstract}

\maketitle

\section{Introduction}

% Measuring quantities (length, electricity...) leads to tolerances and errors
Physical measurement forms the foundation of physics, providing the quantitative data essential for developing theoretical frameworks and predictive models from observations of nature. Every measurement must be accompanied by an uncertainty estimate, reflecting both methodological constraints and instrumental limitations. Consider measuring temperature with a thermometer: uncertainty arises from factors including the thermometer's calibration during manufacturing, the resolution of its smallest markings, and systematic errors occurring during the measurement process. Experimental physics is built upon this fundamental relationship between measurement and uncertainty, requiring that all scientific claims be evaluated within the context of their quantifiable precision limits.

% Same happens with time. Definition of a clock (just citing your paper about the axiomatic principles plus some short intuitive description, nothing too long)
The measurement of time, like the measurement of any physical quantity, requires specialised instruments with finite precision, called clocks \cite{9973001, 804271}. For our analysis of temporal measurement, we define a clock broadly as any physical system that satisfies the definition given in \cite{Woods2021autonomousticking}. More simply, a clock is any device that independently signals the passage of time to its surroundings through a classical signal composed of sequential events---``ticks''---occurring at undetermined moments \cite{silva2023tickingclocksquantumtheory}. Beyond traditional timepieces, this definition encompasses various time-measuring instruments, including timers, stopwatches and metronomes \cite{PhysRevX.11.011046}.

% People want more and more accurate clocks, but there is not an ideal reference frame. In the light of this, we have tried to get as close as possible to a perfect time reference (proof of this is the redefinition of a second in terms of more and more precise oscillators) but never achieved (they require infinite energy; not only in practice but in energy as well due to uncertainty principle)
Our dependence on time measurement spans countless activities, many of which require exceptional precision. Throughout history, we have developed progressively sophisticated timekeeping technologies---a progression clearly illustrated by the evolving definition of the second, which has advanced from measuring pendulum swings to atomic transitions. However, a fundamental challenge persists in how we evaluate timekeeping precision: such evaluation typically requires comparison with either an ideal reference clock or a clock of superior precision to the device being tested \cite{1446564, 1990STIN...9122539S, 804271}. This creates a circular dilemma, as it appears that validating any clock ultimately necessitates comparison with another, more precise timepiece \cite{PhysRevLett.85.2010, Kómár2014}.

% Motivate R (all the theoretical papers on clocks bound R and talk about R.)
It is important to clarify what we mean by a clock's \textit{precision}. While the literature offers various metrics for characterising how precise a clock is, this paper focuses on the parameter $R:=\mu^2/\sigma^2$, where $\mu$ represents the mean time between ticks and $\sigma^2$ is the effective variance per tick of the ticking process \cite{PhysRevX.7.031022, Woods2021autonomousticking, rpls-mp8z}. For reset clocks with independent ticks, $\sigma^2$ reduces to the variance of a single inter-tick interval. For weakly dependent clocks, $\sigma^2$ denotes the asymptotic variance per tick, defined more precisely in \Cref{sec:app_proofs}. This definition is particularly valuable for developing theoretical results for timekeeping devices and applies to both classical and quantum systems \cite{PRXQuantum.3.010319}. Furthermore, as demonstrated in \cite{Dominguez2026MeasuringClockPrecision}, a direct equivalence exists between $R$ and other experimentally defined precision metrics such as Allan variance~\cite{1446564, 5570702, 7406774}. Therefore, given its versatility, utility, and operational nature, we concentrate on estimating $R$ using a single clock throughout the remainder of this work.

In our accompanying work \cite{Dominguez2026MeasuringClockPrecision}, we address this circular dependency in a setting where the only auxiliary resource is another clock of the same precision. There, the central result is that the theoretical precision parameter $R$ is operationally accessible from clock-comparison data alone: an ideal reference, and even a strictly more precise reference clock, is not required. That analysis also clarifies the relation between $R$, double-tick statistics and Allan-type precision measures, thereby establishing the operational meaning of $R$ within a purely clock-based framework.

The present work asks a complementary, stronger question. Rather than using a second clock as a reference object, we ask whether a single clock can characterise its own precision using only elementary signal-processing components that do not themselves constitute a time reference. This change of resource regime is experimentally appealing, since producing a stable delay element can be substantially simpler than producing a second clock with matched precision. At the same time, it introduces a different theoretical problem: the uncertainty of the delay element cannot be ignored, and must be separated from the intrinsic stochasticity of the clock without appealing to an external temporal standard. Thus, while \cite{Dominguez2026MeasuringClockPrecision} establishes the operational status of $R$ through clock-to-clock comparisons, the present paper investigates the more self-referential limit in which the clock is effectively measured against its own tick stream.

%%%%%%%%%%%%%%%%%%%%%%%%%%%%%%%%%%%%%%%%%%%%%%%%%%%%%

\section{Theoretical framework}
\label{sec:theoframe}

% Cite Tomography of clock signals using the simplest possible reference
The central element of our investigation is the clock to be characterised. In accordance with the definition proposed in \cite{Woods2021autonomousticking}, we conceptualise clocks as physical systems that autonomously transmit temporal information to their environment through discrete ticks. Given the absence of an ideal time reference frame, the time interval between consecutive ticks must necessarily be treated not as a deterministic quantity but as a random variable $T$.

While the most general definition of a clock does not impose constraints on the properties of these random variables, our work focuses specifically on clocks whose ticks are weakly dependent (see \Cref{sec:app_proofs} for a rigorous definition). Broadly speaking, this family of clocks includes those whose current ticking behaviour is mainly governed by the last few ticks, with diminishing influence from ticks further in the past. Recent feedback-controlled ticking-clock models provide a concrete mechanism for this kind of history dependence: classical information extracted from previous ticks can modulate the subsequent clockwork dynamics while preserving the autonomous ticking-clock structure \cite{Miller2026FeedbackClockworks}. This model appears well suited to representing some currently proposed implementations of quantum clocks \cite{Culhane_2024}. Reset clocks, which return to their initial state after producing each tick, are also included here: their ticks occur independently of previous ones and therefore follow identical probability distributions, making the ticking times $T_i$ independent and identically distributed (i.i.d.) random variables \cite{Woods2021autonomousticking, PhysRevX.6.041053}.

% Picture with single clocks and ticks in time line
\begin{figure}[htbp]
    \centering
    \includegraphics[width=0.5\textwidth]{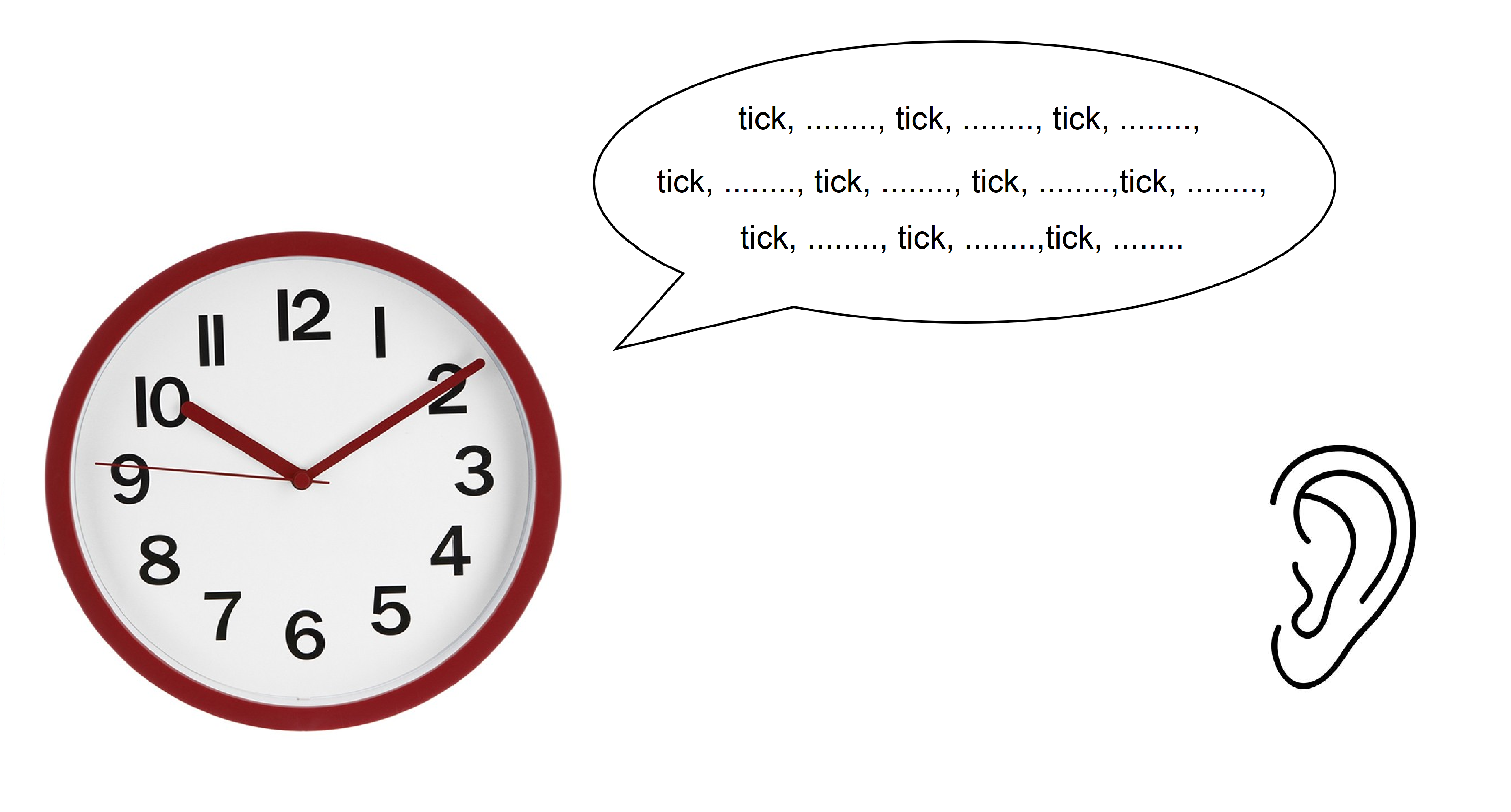}
    \caption{Normal operation of a clock: a physical system capable of transmitting information to the environment about the passage of time in the form of autonomous ``ticks''.}
    \label{fig:ticks}
\end{figure}

% Definition of a delay trigger (as formal as possible)
The second element of the experiment is the delayed trigger. Unlike clocks, this type of device currently has no precise definition in the literature, so we introduce one here. 

A \textit{delayed trigger} is a system with a single input $s_{in}(t)$ and a single output $s_{out}(t)$, such that:

\begin{equation}
    \label{eq:defDelay}
    s_{out}(t) =
    \begin{cases}
        0 & \text{if } t < D\ \text{ or }\ \max\limits_{0 \leq \tau \leq t-D} \{s_{in}(\tau)\} \leq \frac{1}{2} \\
        1 & \text{otherwise}
    \end{cases}.
\end{equation}

Delayed triggers are characterised by their delay parameter $D$. We use a delayed trigger instead of a delay line because, theoretically, the delayed trigger is a less complex system with more elementary behaviour, thereby avoiding possible loopholes through which the device could be exploited to construct a new clock or another timekeeping device.

Operationally, one can find some similarities between the delayed trigger from this experiment and the second clock from our accompanying paper \cite{Dominguez2026MeasuringClockPrecision}: the delayed trigger replaces the second clock by supplying the stopping event against which the ticks of the clock under test are counted. Repeating the experiment produces a distribution of tick counts whose variance contains two contributions: the accumulated fluctuations of the clock and the trial-to-trial fluctuations of the trigger delay. Once the latter contribution is independently bounded or estimated, the former determines $R$. The delayed trigger therefore plays the limited role needed from the second clock, marking the end of each observation interval, without generating a periodic tick stream or defining a frequency standard of its own.

In practical experimental implementations, factors such as ambient temperature fluctuations and environmental interactions may introduce slight variations in the delay parameter $D$ across successive experimental iterations. Furthermore, the absence of an ideal time reference frame renders precise determination of $D$ with zero uncertainty fundamentally impossible. Consequently, in the subsequent analysis, $D$ is treated as a random variable rather than a deterministic parameter. The sole assumption imposed on this variable is that its relative fluctuations remain small (a reasonable constraint, since extended delays with minimal relative uncertainty are readily achievable in experimental practice). Using fibre-spool delay lines is a common practice in experimental photonics \cite{Wang, K_f_lian_2009, Volyanskiy, hilton2019heterodynefiberinterferometerfrequencynoise}.

For methodological completeness, this work incorporates a third device: a signal splitter. Without loss of generality, we characterise this component as a single-input, dual-output device in which both outputs replicate the input signal after experiencing identical propagation delays and amplitude attenuation. This symmetric behaviour in both delay and attenuation characteristics is maintained across the two output channels.

%%%%%%%%%%%%%%%%%%%%%%%%%%%%%%%%%%%%%%%%%%%%%%%%%%%%%%%%%%

\section{Theoretical results}
\label{sec:results}

\begin{figure}[htbp]
    \centering
    \includegraphics[width=0.48\textwidth]{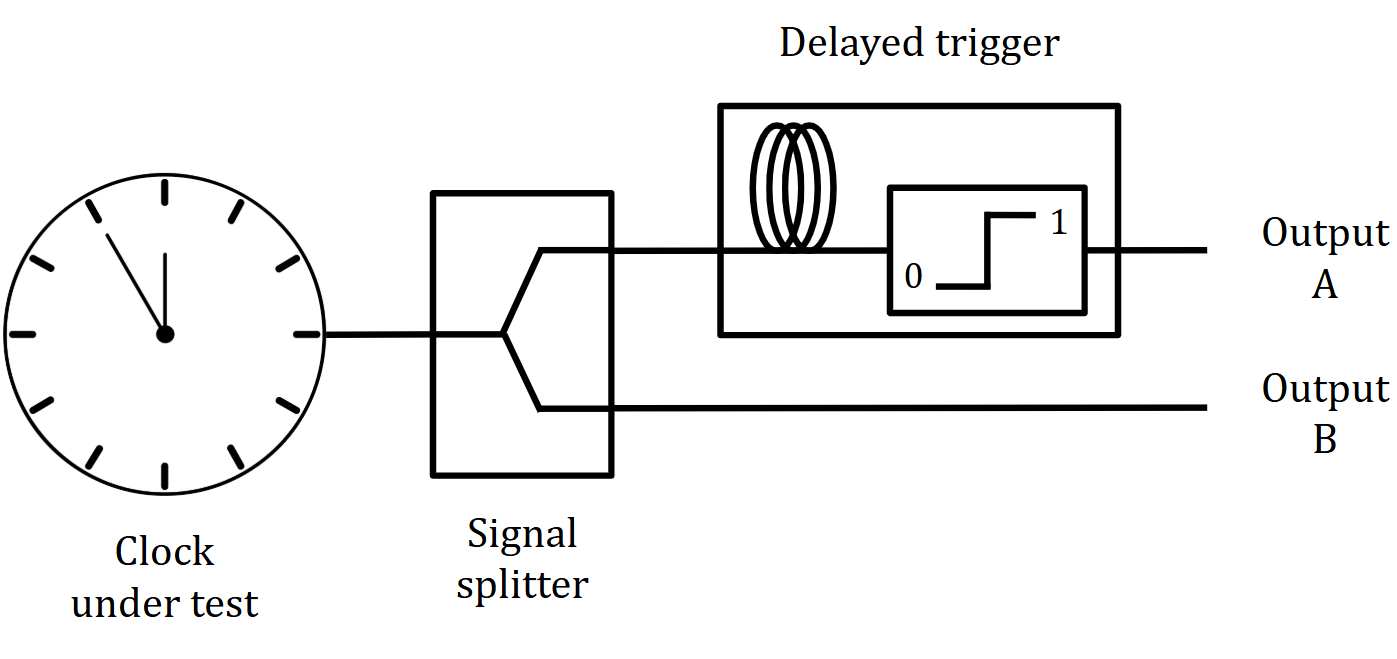}
    \caption{Block diagram depicting the proposed setup. The clock whose precision is to be measured is connected to a signal splitter, and one of the splitter outputs is used as an input for a delayed trigger. During the experiment, the user counts the ticks received at output B before the signal arrives at output A.}
    \label{fig:blocks}
\end{figure}

The main result is a method for characterising the precision of a clock using only the three elements described in the previous section: the clock to be characterised, a delayed trigger and a signal splitter. These must be connected as described in \Cref{fig:blocks}.

\Cref{th:1} then holds:

\begin{restatable}{theorem}{maintheorem}
\label{th:1}

Consider the clock described in~\Cref{fig:blocks}. Then, 

\begin{enumerate}
    \item Turn the clock on at an arbitrary time labelled $t=0$.
    \item Count the number of ticks received at output B, that is, the output that is not connected to the delayed trigger.
    \item As soon as output A (the one coming from the delayed trigger) switches from 0 to 1, stop the count, note the number of ticks recorded (which we denote by $\tau_D$) and reset the experiment.
    \item Repeat steps 1 - 3 for the number of iterations required to achieve the desired precision in the estimates of $\mathbb{E}[\tau_D]$ and $\Delta^2(\tau_D)$.
\end{enumerate}

In the asymptotic regime $D \to \infty$ with $\Delta^2(D)/\mu^2=o(\mathbb{E}[\tau_D])$, the precision of the clock $R$ is given by:

\begin{equation}
\label{eq:res_1}
    R = \frac{\mathbb{E}[\tau_D]}{\Delta^2(\tau_D)-\frac{\Delta^2(D)}{\mu^2}}.
\end{equation}

Here $\Delta^2(D)$ is the variance of the delay parameter in the delayed trigger, which need not be constant across iterations of the experiment, and $\mu$ is the average ticking period of the clock at hand.
\end{restatable}

The proof of \cref{eq:res_1}, detailed in \Cref{sec:app_proofs}, uses formulations of the central limit theorem for weakly dependent stochastic processes, as well as results from renewal theory \cite{libroProcesos}. Furthermore, \Cref{sec:app_proofs} gives the leading finite-$D$ expression for $R$ under the same asymptotic assumptions.

\Cref{eq:res_1} is particularly useful in scenarios where $\Delta^2(\tau_D) \gg \Delta^2(D)/\mu^2$, namely when the clock being characterised has significantly lower precision than the delayed trigger. Under these conditions, since both $\mathbb{E}[\tau_D]$ and $\Delta^2(\tau_D)$ can be measured directly, the precision parameter $R$ can be determined with high accuracy. This assumption is physically justified, as fabricating a high-precision delayed trigger presents substantially fewer technical challenges than developing a precise clock, assuming access to the requisite manufacturing capabilities.

Nevertheless, the question of whether it is possible to obtain $R$ in setups where $\Delta^2(\tau_D) \approx \Delta^2(D)/\mu^2$ without relying on an external reference is also important. Within the constraints of the original experimental design (consisting of a single clock and a single delayed trigger), the independent operational measurement of $\sigma$ and $\Delta^2(D)/\mu^2$ appears infeasible, as $\Delta^2(\tau_D)$ will invariably incorporate both uncertainty components. However, if the clock's ticking frequency can be systematically adjusted without compromising its precision (effectively scaling both $\mu$ and $\sigma$ by an identical constant), then \Cref{th:2} can be used to measure the delayed-trigger uncertainty operationally:

\begin{restatable}{theorem}{maintheoremtwo}
\label{th:2}

Perform the experiment described in \Cref{th:1} $n\ge3$ times, tuning the ticking frequency of the clock to a different value between experimental runs in a way that does not alter its precision $R$. Then, in the asymptotic regime of \Cref{th:1}, the normalised uncertainty in the final measurement due to the delayed trigger can be obtained by

\begin{equation}
\label{eq:res_2}
    \frac{\Delta^2(D)}{\mu^2_{(j)}} =
    \frac{\mathbb{E}[\tau_D^{(i)}]\Delta^2(\tau_D^{(j)}) - \mathbb{E}[\tau_D^{(j)}]\Delta^2(\tau_D^{(i)})}
    {\mathbb{E}[\tau_D^{(i)}]-\frac{\mu^2_{(j)}}{\mu^2_{(i)}}\mathbb{E}[\tau_D^{(j)}]}
    + o(1).
\end{equation}

Here $i,j \in \{1, 2, \ldots, n\}$ denote the indices of the frequency settings used.
\end{restatable}

All variables on the right-hand side of \cref{eq:res_2} can be measured directly except for the quotient $\mu^2_{(j)}/\mu^2_{(i)}$ in the denominator. This quotient can be obtained without relying on an external reference frame by repeating the experiment with at least three different ticking-frequency settings and fitting the common-delay scaling of the mean tick counts. Substituting the resulting estimate of $\Delta^2(D)/\mu^2$ into \cref{eq:res_1} gives an operational estimate of $R$ that depends only on measured quantities.

Alternative measurement strategies do not improve our knowledge of $\Delta^2(D)/\mu^2$. For instance, implementing a modified tick-counting protocol (e.g., counting every $n$ ticks) does not yield additional precision information. While such a method would create a new clock with precision $nR$ (where $\mu^{(n)} = n\mu$ and $\sigma^{(n)} = \sqrt{n}\sigma$), the right-hand side of \cref{eq:res_1} would correspondingly scale by a factor of $n$, resulting in no net informational gain. Similarly, employing multiple delayed triggers with distinct delay parameters $D$ proves problematic, as characterising the uncertainty in each delay parameter would necessitate an external time reference---precisely what we aim to avoid in this measurement framework.

%%%%%%%%%%%%%%%%%%%%%%%%%%%%%%%%%%%%%%%%%%%%%%%%%%%%%%%%%%%%%%%%%%%%%%

\section{Experimental implementation: the clock that measures its own precision}

\iffalse
EXP 1:
=== Temperature corrected Data ===
Mean: 168021.6505 ± 0.6822
Measured Variance: 138.6255
Quantization Variance: 21.3333
Real Variance: 117.2922 ± 10.4700
Real Std: 10.8302 ± 0.4834
R (mean/variance): 1432.504852 ± 127.871641

EXP 2:
=== Temperature corrected Data ===
Mean: 18408.4557 ± 0.1809
Measured Variance: 73.3512
Quantization Variance: 21.3333
Real Variance: 52.0178 ± 1.8455
Real Std: 7.2123 ± 0.1279
R (mean/variance): 353.887507 ± 12.555046
\fi

\begin{figure}[htbp]
    \centering
    \includegraphics[width=0.48\textwidth]{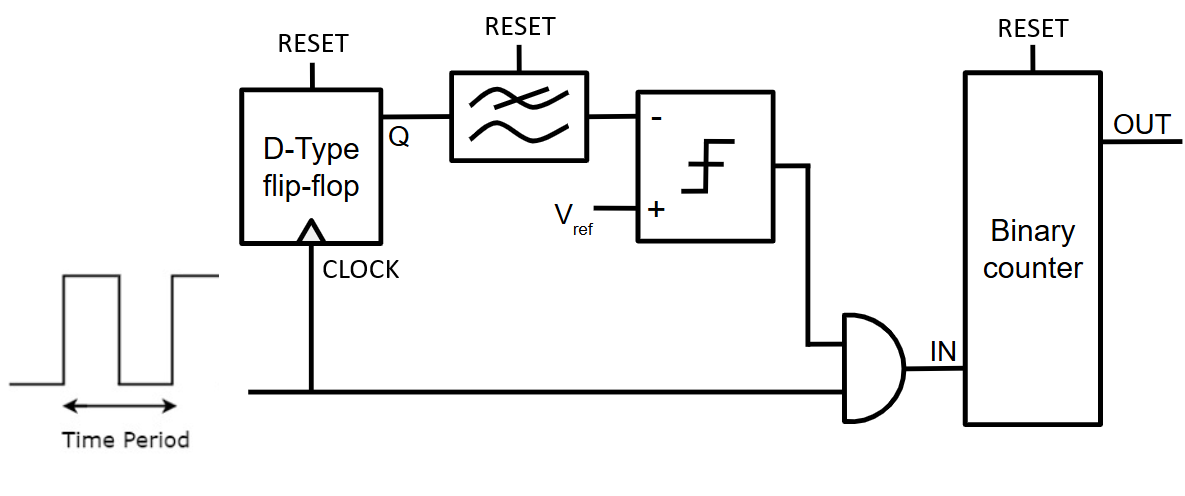}
    \caption{Block diagram depicting the realised experiment. The clock signal is split into two branches, one of which passes through the delayed trigger (implemented by a D-type flip-flop feeding an RC low-pass filter, whose output is rectified with a voltage comparator) and meets the other branch at an AND gate. The ticks pass through the gate until the signal reaches the output of the delayed trigger, and they are counted in the binary counter. After the microcontroller reads out the number of ticks, the experiment is reset via the RESET signal.}
    \label{fig:exp_blocks}
\end{figure}

To test the methods described in \Cref{sec:results} for characterising a clock, we designed and built an electronic implementation following the block diagram in \Cref{fig:exp_blocks}. The design files and firmware can be found in \cite{alvarez_dominguez_2026_22739518}.

The measurement board comprises an input channel for the clock signal under test, a digital output for the registered tick count, a $5$ V power supply input, a reset input for reinitialising the system between measurement cycles, and a PT1000 thermistor for precise temperature monitoring.

The circuit operates as follows. The incoming clock signal must have adequate amplitude but need not be a square wave. It is initially divided into two parallel signal paths, the first of which feeds into the delayed trigger circuit.

The delayed trigger consists of four cascaded stages. In the first stage, a D-type flip-flop with its data input tied to logic high is clocked by the incoming signal. Upon arrival of the first rising edge, the flip-flop transitions from logic low to logic high and maintains this state until externally reset. The propagation delay of this stage contributes to the overall timing uncertainty of the delayed trigger. The second stage comprises an RC filter that transforms the abrupt logic transition from the flip-flop into a gradually ramping voltage, thereby generating the desired time delay. The delay parameter $D$ and its associated uncertainty $\Delta D$ are primarily determined by the RC time constant of this stage. In the third stage, a voltage comparator with a fixed reference voltage on one input and the filtered ramp signal on the other converts the slowly varying voltage into a sharp logic transition once the threshold is exceeded. The fourth stage consists of an inverting buffer (NOT gate), configured so that the delayed trigger outputs logic high before the delayed activation and logic low thereafter.

The delayed trigger output is combined with the second signal path from the initial splitter via an AND gate, which transmits clock ticks only while the delayed trigger remains inactive. A binary counter accumulates the transmitted ticks and presents the count via a parallel output port.

The measurement procedure is straightforward: each measurement cycle begins by pulsing the reset input to clear the counter, reset the flip-flop and discharge the capacitor. Clock ticks are then applied to the input, and the counter output is read after the delayed trigger has activated, yielding the tick count for that iteration. The system is then reset for subsequent measurements.

As shown in \cref{eq:res_1}, accurate characterisation of $R$ requires minimising the ratio $\Delta^2(D)/\mu^2$. The delay $D$ between experimental trials is influenced by several factors, principally temperature fluctuations, power supply voltage ripple, and component ageing. In our case, component ageing can be neglected, as the experimental timescales are orders of magnitude shorter than those associated with changes in circuit characteristics. Power supply ripple can be suppressed to arbitrarily low levels through appropriate decoupling capacitors or low-pass filters placed adjacent to sensitive components, combined with the use of high-stability laboratory power supplies.

Temperature fluctuations present a more significant challenge. Optimal thermal stability is achieved using an array of metallised polypropylene (MPP) capacitors connected in parallel. This capacitor technology exhibits the lowest thermal variability, with a linear temperature coefficient of approximately 40 ppm/°C, and the parallel configuration provides additional thermal averaging. For the resistive elements, bulk metal foil or precision metal film resistors are preferred, both offering temperature coefficients below 1 ppm/°C. The resistance should be minimised within practical constraints to reduce thermal noise contributions, and values in the range of 100 k$\Omega$ to 1 M$\Omega$ represent an optimal compromise.

\begin{figure}[htbp]
    \centering
    \includegraphics[width=0.48\textwidth]{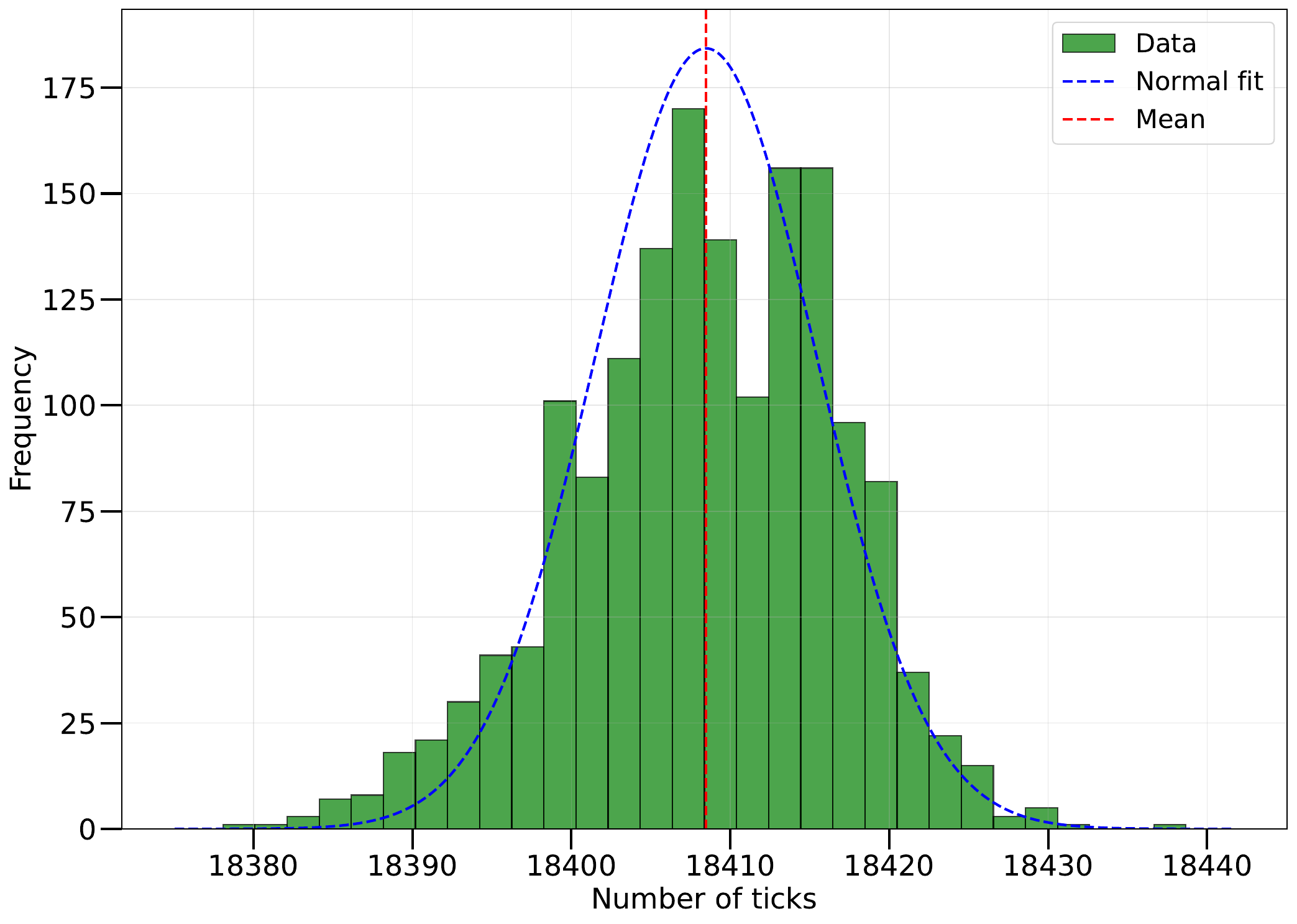}
    \caption{Experimental results. After $1600$ iterations and the corresponding temperature corrections, the delayed trigger exhibited an estimated delay of $18408.46 \pm 0.18$ ticks. The clock under test produced a standard deviation in the number of ticks counted of $7.21 \pm 0.12$, which corresponds to a measured precision $R = 353.89 \pm 12.5$. The error in the measurement of $R$ can be reduced arbitrarily by increasing the number of iterations performed.}
    \label{fig:exp}
\end{figure}

\Cref{fig:exp} presents the experimental results. The test clock was generated by an ATmega 328P microcontroller driven by a temperature-compensated crystal oscillator (TCXO); one output pin was toggled between $0$ V and $5$ V at a frequency of $2^{12}$ Hz. A total of $1600$ iterations were performed, providing an optimal compromise between measurement precision and data acquisition time. Because the delay time $D$ is temperature dependent, corrections were applied based on temperature measurements from the integrated PT1000 thermistor to ensure temporal consistency of the acquired data. The resulting distribution exhibits excellent agreement with a normal distribution, as predicted by \Cref{sec:results}. The mean tick count for this configuration was $18408.46 \pm 0.18$, with a standard deviation of $7.21 \pm 0.12$, yielding $R = 353.89 \pm 12.5$.

To validate this result, we calculated the expected standard deviation of the clock-generated ticks via conventional error propagation. Since the output pin state is toggled by polling the device's internal microsecond counter (updated every $4$ $\mu$s), the generated ticks exhibit an intrinsic period uncertainty of $4/\sqrt{12} \approx 1.15$ $\mu$s \cite{arduino}. Concurrent readout operations from the binary counter (required to handle overflows) contribute a further uncertainty of $5$ to $7$ $\mu$s, as these operations have variable duration depending on internal triggers and other processes detailed in the microcontroller data sheet \cite{atmega328p}. This yields a net standard deviation in the tick period between $5.13$ $\mu$s and $7.09$ $\mu$s, corresponding to values of $296.43 \le R \le 566.22$. These values are in agreement with the experimentally measured value.

%%%%%%%%%%%%%%%%%%%%%%%%%%%%%%%%%%%%%%%%%%%%%%%%%%%%%%%%%%%%%%%%%%%%%%%%%%%

\section{Discussion and outlook}
\label{sec:discussion}

The methodology presented in this paper represents a significant advance in metrology, particularly for time measurement. By enabling the characterisation of a clock's precision without requiring any external time reference, we address a fundamental circular dependency in timekeeping technology. This self-referential approach has both theoretical significance and practical applications.

The primary strength of our method lies in its ability to provide a reliable estimate of a clock's precision parameter $R$ through the analysis of the clock's own ticking behaviour. This is particularly valuable when characterising extremely precise clocks, where finding a more accurate reference timepiece would traditionally present significant technical challenges. Our approach circumvents this limitation by using only the statistical properties of the clock's output and a delayed trigger system.

However, several practical considerations should be noted. Convergence to a precise estimate of $R$ may be slow, especially when the clock under test exhibits high precision relative to the delayed trigger (i.e., when $\Delta^2(\tau_D)$ approaches $\Delta^2(D)/\mu^2$). In such scenarios, a large number of experimental iterations may be required to achieve statistically precise results. To enhance convergence speed, one could perform multiple experimental runs using different delayed triggers with varying delay parameters $D$. This approach would provide a more diverse dataset and potentially enable more efficient estimation of $R$.

A significant open question remains regarding whether one can accurately measure the variance of the delay parameter ($\Delta^2(D)/\mu^2$) without modifying the clock under test. While \Cref{th:2} offers an operational approach when the clock's ticking frequency can be systematically adjusted, this assumes that such adjustments do not affect the underlying precision of the timepiece. Further theoretical and experimental work is needed to determine whether alternative methodologies could enable direct measurement of $\Delta^2(D)/\mu^2$ without such frequency tuning.

This self-referential precision measurement technique may find applications in fields beyond fundamental physics. In engineering contexts, it could enable more robust autonomous systems where external calibration is impractical or impossible. The method might also prove valuable in standardisation efforts for timekeeping, potentially offering new approaches to characterising precision in timing applications ranging from telecommunications to financial transaction systems.

Further theoretical development could establish the ultimate boundaries of self-referential precision assessment, helping to determine where this approach offers advantages over conventional methods and where its limitations might necessitate alternative solutions.

%%%%%%%%%%%%%%%%%%%%%%%%%%%%%%%%%%%%%%%%%%%%%%%%%%%%%%%%%%%%%%%%%%%%%%%%%%%%%%%%%%%%

\section{Conclusion}

In this work, we have presented a methodology for characterising the precision of a timekeeping device without relying on any external time reference. By analysing the statistical properties of a clock's own ticking behaviour in conjunction with a delayed trigger system, we have derived a relationship that allows the precision parameter $R = \mu^2/\sigma^2$ to be determined through purely self-referential measurements.

Our main result demonstrates that, under reasonable physical assumptions, the precision $R$ can be calculated simply by counting the ticks produced by the clock and computing the first two statistical moments, using quantities that can be experimentally measured without external time references. This approach is particularly effective when the clock under test has significantly lower precision than the delayed trigger component.

To validate our theoretical framework, we developed a specialised electronic circuit implementing the proposed methodology. In the experiment reported here, the test clock was generated by an ATmega 328P microcontroller running from a TCXO clock source. The empirical results confirmed our analytical predictions, demonstrating that a clock's precision can indeed be characterised accurately without referencing any external timekeeper.

The practical implementation of our self-referential methodology not only confirms its theoretical validity but also opens new possibilities for autonomous calibration in isolated systems. This approach resolves a fundamental circularity in metrological practice and enables precision measurements in contexts where external references are impractical or undesirable, such as space missions, autonomous quantum systems, or fundamental physics experiments where reference frames might introduce unwanted artefacts.

\begin{acknowledgments}
This work was supported by the MSCA Cofund QuanG (Grant Number: 101081458), funded by the European Union. The views and opinions expressed are those of the author(s) only and do not necessarily reflect those of the European Union. Neither the European Union nor the granting authority can be held responsible for them.

This manuscript is intended to be read alongside the companion work \cite{Dominguez2026MeasuringClockPrecision}, which studies the complementary problem of clock characterisation using equally precise clock copies rather than a single-clock delayed-trigger setup.

The schematics, PCB layouts, firmware and readout code are available in the Zenodo repository \cite{alvarez_dominguez_2026_22739518}.

None of the results in this manuscript were derived in an AI-assisted manner. The only use of AI was for typesetting and formatting the final manuscript. Claude, Qwen and ChatGPT were used for this task.

\begin{comment}
\begin{figure}[htbp]
    \centering
    \begin{minipage}{0.12\textwidth}
        \centering
        \includegraphics[width=\textwidth]{img_files/imagen_EU.jpg}
    \end{minipage}
    \hspace*{0.08\textwidth} % Adds horizontal space between images
    \begin{minipage}{0.12\textwidth}
        \centering
        \includegraphics[width=\textwidth]{img_files/logo-quang.png}
    \end{minipage}
\end{figure}
\end{comment}

\end{acknowledgments}

\appendix
\crefname{appendix}{appendix}{appendices}
\Crefname{appendix}{Appendix}{Appendices}
\crefalias{section}{appendix}
\renewcommand{\theHequation}{\theHsection.\arabic{equation}}
\section{Proof of the main results}
\label{sec:app_proofs}

\subsection{Proof of the first theorem}

Let $T_i$ be the duration of the $i$-th tick, as defined in the main text, and let $\bar T_i := T_i-\mu$ be its centred version. Our results apply to clocks whose stochastic process $\{\bar T_i\}_{i\in\mathbb{Z}}$ is stationary and satisfies conditions (i), (ii) and (iii) of Assumption 2.1 in \cite{CLT}, for some $p>3$.

Condition (i) ensures that the random variables have a finite $p$-th absolute moment norm, $\mathbb{E}[|\bar T_i|^p]^{1/p}<\infty$. Condition (ii) describes how strongly the current zero-mean tick duration $\bar T_i$ is affected by the remote past: physically, disturbances introduced far in the past have a diminishing influence on the present, with sufficiently fast polynomial decay. Condition (iii) is a non-triviality condition asserting that the summed process exhibits persistent fluctuations. Under these assumptions, define the effective variance per tick by
\begin{equation}
\label{eq:app_sigma_eff}
    \sigma^2 :=
    \lim_{n\to\infty}\frac{1}{n}\operatorname{Var}\!\left(\sum_{i=1}^n \bar T_i\right).
\end{equation}
When the covariance-series representation is valid, this can equivalently be written as
\begin{equation}
\label{eq:app_sigma_cov}
    \sigma^2 =
    \sum_{\ell\in\mathbb{Z}}\mathbb{E}[\bar T_0\bar T_\ell].
\end{equation}
The precision parameter used in the main text is then $R:=\mu^2/\sigma^2$. In the specific case of reset clocks, the ticking times $T_i$ are i.i.d.; therefore, condition (ii) is satisfied by definition, condition (iii) holds whenever $\operatorname{Var}(T_i)>0$, and \cref{eq:app_sigma_eff} reduces to the usual single-tick variance.

With these definitions, let $D$ be the delay parameter of the delayed trigger, and define the tick-counting function
\begin{equation}
\label{eq:app_A1}
    \tau_t = \sup\biggl\{n\ :\sum_{i=1}^n T_i \le t\biggr\}.
\end{equation}
The delay $D$ is treated as a random variable, since environmental fluctuations can vary it between experimental runs. Similarly, $\tau_t$ is random because the number of ticks produced within a given physical duration is not deterministic. Finally, define $S_n:=\sum_{i=1}^nT_i$.

Conditioning on a particular value of $D$, we have
\begin{equation}
\label{eq:app_A2}
    \mathbb{E}[\tau_D|D]
    = \sum_{n=1}^\infty P(S_n<D).
\end{equation}
Likewise,
\begin{align}
\label{eq:app_A3}
    \mathbb{E}[\tau_D^2|D]
    &= \sum_{n=1}^\infty (2n-1)P(S_n<D).
\end{align}
By Theorem 2.4 of \cite{CLT}, the centred cumulative sums admit a normal approximation in the central window $|n-D/\mu|\le C\sqrt{D}$. Thus the cumulative sums, not the individual tick increments, are approximated by Gaussian variables in the large-$D$ regime:
\begin{align}
\label{eq:app_A4_def}
    \mathbb{E}[\tau_D|D]
    &= \sum_{n=1}^\infty\Phi\left(\frac{D-n\mu}{\sigma\sqrt{n}}\right)+o(1),\\
\label{eq:app_A4_def2}
    \mathbb{E}[\tau_D^2|D]
    &= \sum_{n=1}^\infty(2n-1)\Phi\left(\frac{D-n\mu}{\sigma\sqrt{n}}\right)+o(1),
\end{align}
where $\Phi(x)$ denotes the cumulative distribution function of the normal distribution. The approximation error is uniform in the central window, while the complementary tails are asymptotically negligible, so summing over $n$ preserves the little-$o$ order as $D\to\infty$.

Applying the standard renewal/random-walk asymptotics to this Gaussian approximation \cite{RenewalTheory} gives
\begin{equation}
\label{eq:app_A4}
    \mathbb{E}[\tau_D|D]
    = \frac{D}{\mu} + \frac{1}{2R} - \frac{1}{2} + o(1),
\end{equation}
and
\begin{equation}
\label{eq:app_A5}
    \Delta^2(\tau_D|D)
    = \frac{D}{R\mu} + \frac{5}{4R^2} + \frac{1}{12} + o(1).
\end{equation}
Assume now that $D$ is independent of the ticking process and has finite moments compatible with the preceding asymptotic expansions. The laws of total expectation and total variance then yield
\begin{align}
\label{eq:app_A7}
    \mathbb{E}[\tau_D]
    &= \frac{\mathbb{E}[D]}{\mu} + \frac{1}{2R} - \frac{1}{2} + o(1),\\
\label{eq:app_A7_var}
    \Delta^2(\tau_D)
    &= \frac{\mathbb{E}[D]}{R\mu} + \frac{5}{4R^2} + \frac{1}{12}
    + \frac{\Delta^2(D)}{\mu^2} + o(1).
\end{align}

Solving \cref{eq:app_A7,eq:app_A7_var} for $R$ gives the leading finite-$D$ expression
\begin{widetext}
\begin{equation}
\label{eq:app_A8}
    R =
    \frac{
        6\left(\mathbb{E}[\tau_D]
        +\sqrt{\mathbb{E}[\tau_D]^2+\mathbb{E}[\tau_D]
        +3\Delta^2(\tau_D)-3\frac{\Delta^2(D)}{\mu^2}}\right)+3
    }{
        12\Delta^2(\tau_D)-12\frac{\Delta^2(D)}{\mu^2}-1
    }
    +o(1).
\end{equation}
\end{widetext}
For large $D$, and assuming that $\mathbb{E}[\tau_D]$ dominates both the finite-$D$ correction terms and $\Delta^2(D)/\mu^2$, \cref{eq:app_A8} reduces to \cref{eq:res_1}.

\qed

\subsection{Proof of the second theorem}

For each tuned ticking-frequency setting $k$, denote
\begin{equation}
\label{eq:app_B1}
    m_k:=\mathbb{E}[\tau_D^{(k)}],
    \qquad
    v_k:=\Delta^2(\tau_D^{(k)}),
    \qquad
    d_k:=\frac{\Delta^2(D)}{\mu_{(k)}^2}.
\end{equation}
The frequency tuning is assumed to scale both $\mu$ and $\sigma$ by the same factor, so the precision $R=\mu^2/\sigma^2$ is unchanged. Applying \cref{eq:res_1} at each frequency gives, in the same asymptotic regime,
\begin{equation}
\label{eq:app_B2}
    v_k = \frac{m_k}{R}+d_k+o(1).
\end{equation}
For two settings $i$ and $j$, write
\begin{equation}
\label{eq:app_B3}
    q_{ij}:=\frac{\mu_{(j)}^2}{\mu_{(i)}^2},
    \qquad
    d_i=q_{ij}d_j.
\end{equation}
Substituting this relation into \cref{eq:app_B2} for $i$ and $j$ gives
\begin{align}
\label{eq:app_B4}
    v_i &= \frac{m_i}{R}+q_{ij}d_j+o(1),\\
    v_j &= \frac{m_j}{R}+d_j+o(1).
\end{align}
Eliminating $1/R$ between the two equations yields
\begin{equation}
\label{eq:app_B5}
    d_j =
    \frac{m_i v_j-m_j v_i}{m_i-q_{ij}m_j}
    +o(1).
\end{equation}
Substituting the definitions of $m_k$, $v_k$, $d_j$ and $q_{ij}$ back into \cref{eq:app_B5} recovers \cref{eq:res_2}.

It remains to note that no external time reference is required to determine the dimensionless ratio $q_{ij}$. At leading order in \cref{eq:app_A7}, the mean tick counts scale with the common physical delay as $m_k=\mathbb{E}[D]/\mu_{(k)}+O(1)$. Repeating the experiment with at least three ticking-frequency settings therefore allows the common-delay scaling and the dimensionless period ratios to be fitted from tick-count data alone. Once $d_j$ has been obtained from \cref{eq:app_B5}, substituting it into
\begin{equation}
\label{eq:app_B6}
    R = \frac{m_j}{v_j-d_j}+o(1)
\end{equation}
gives an operational estimate of the clock's precision that depends only on counted ticks and dimensionless frequency ratios.

\qed

\nocite{*}
\bibliography{apssamp}% 

\end{document}